%% file: ms.tex
\documentclass[12pt,preprint]{aastex}
\usepackage{graphicx}

\newcommand{\water}{H$_{2}$O~}
\newcommand{\myemail}{mcgrath@ifa.hawaii.edu}
\newcommand{\asec}{$^{\prime\prime}$}   
\newcommand{\kms}{km~s$^{-1}$}      
\newcommand{\1}{\phantom{.0}}      
\newcommand{\0}{\phantom{0}}      
\newcommand{\ras}{\mbox{$.\!\!^{s}$}}

\shorttitle{\water Masers in W49N and Sgr B2}
\shortauthors{McGrath, Goss, & DePree}

\begin{document}

\title{\water Masers in W49 North and Sagittarius B2}

\author{Elizabeth J. McGrath\altaffilmark{1}}
\affil{Institute for Astronomy, University of Hawaii, Honolulu, HI 96822}
\email{\myemail}

\author{W. M. Goss}
\affil{National Radio Astronomy Observatory, Soccoro, NM 87801}
\email{mgoss@nrao.edu}

\and

\author{Christopher G. DePree}
\affil{Department of Physics and Astronomy, Agnes Scott College, Decatur, GA 30030}
\email{cdepree@agnesscott.edu}

\altaffiltext{1}{NSF sponsored REU student, National Radio Astronomy 
	Observatory}
%\altaffiltext{2}{Previously at Vassar College, Physics and Astronomy, 
%	Poughkeepsie, NY 12604}

\begin{abstract}
Using the Very Large Array (VLA) of the National Radio Astronomy 
Observatory in the A and B configurations, we have obtained simultaneous high 
resolution observations of both the 22 GHz \water maser lines as well as the
22 GHz continuum for the H~{\sc ii} regions W49N and Sagittarius B2. The angular
resolution of both observations is $\sim$ 0\farcs1, which at the distance of W49N (11.4 kpc; Gwinn, Moran, \& Reid 1992) and Sgr B2 (8.5 kpc) corresponds to a physical size of $<$1000 AU in both sources.
The velocity coverage
for W49N is $\pm$435 \kms; positions for 316 \water maser components were 
obtained. The velocity coverage for Sgr B2 is -40 to +120 \kms; positions
for 68 maser components were determined in Sgr B2 Main, 79 in Sgr B2 North 
,
14 in Sgr B2 Mid-North, and 17 in Sgr B2 South, for a total of 178 \water maser positions in Sgr B2.

The cross calibration scheme of Reid \& Menten (1990, 1997)
was used.  
Using this procedure, high dynamic range continuum images were obtained with accurate registration ($\sigma\sim$0\farcs01) of the continuum and maser positions.
A detailed comparison between H~{\sc ii} components and maser positions for both Sgr B2 and W49N is presented.  In Sgr B2 Main, the \water masers are predominantly located at the outside edge of the high-frequency continuum, lending support to the proposal that entrainment by stellar winds may play an important role in \water maser emission.
\end{abstract}


\section{Introduction}
\water masers have long been known to be signposts of active star formation and regions of extremely dense ($>10^6$ cm$^{-3}$) gas.  As such, they have proven to be very useful tools to study the dynamics and kinematics of the environment around young, massive stars,
 tracing 
high velocity outflows and shock fronts \citep{eli95}.
Additionally, masers are extremely short-lived phenomena, and therefore pinpoint a precise period in a star's evolution.  In particular, \water masers are often associated with ultra-compact (UC) H~{\sc ii} regions, the short-lived phase that can be associated with massive stars as they move towards the main sequence and produce UV photons which ionize the nearby gas (eg., Garay \& Lizano 1999).  

These observations focus on two well-known high mass star forming regions in our galaxy, W49A and Sgr B2.  W49A has been studied extensively at radio wavelengths (eg., Mezger, Schraml, \& Terzian 1967; Moran et al. 1973; Walker, Matsakis, \& Garcia-Barreto 1982; Gwinn, Moran \& Reid, 1992; DePree et al., 2000).  \citet{mez67} observed W49A to have two thermal
components: A2, 
a small
(diameter = 1 pc) high-density (n$_e \sim$10$^{4}$ cm$^{-3}$) source embedded in a
second source, A1, which is larger (diameter = 14 pc) and lower density (n$_e \sim$
200 cm$^{-3}$). In their investigation, \citet{mez67} pointed out that the
high-density sources might surround O stars still embedded in their
ionized dust cocoons.  
W49A has also been the subject of several investigations at infrared (IR)
wavelengths. 
\citet{con02} have carried out
near-infrared observations of the W49A region that indicate the presence
of approximately 100 O stars. \citet{alv03} have also observed this region in the near-IR, and their J, H, and
K band images indicate the presence of four star-forming clusters
in W49A.  

Located within W49A, the source W49N is composed of several compact and 
ultra-compact H~{\sc ii} regions in a ring formation spanning $\sim$20$^{\prime\prime}$, or 1pc in diameter (e.g., DePree et al.~2000). 
Fig.~\ref{fig.w49} shows the 3.6 cm continuum image of the entire W49A complex with the H~{\sc ii} regions where maser activity is prominent (W49N) indicated.  These regions are shown in more detail in the inset 1.3 cm continuum images.  The H~{\sc ii} region G1/G2 is the main locus of water masers in W49N, along with a smaller cluster of \water masers located near the H~{\sc ii} region B1/B2.  The \water 
maser velocity distribution in W49N is known to span at least 560 \kms~
\citep{wal82}, and is the most luminous source of \water maser activity in the Galaxy.

The Sgr B2 star-forming region and its ultra- and hyper-compact H~{\sc ii} regions have also been studied in great detail (eg., Gaume et al. 1995; DePree et al. 1995, 1996; DePree, Goss \& Gaume 1998).  A variety of morphologies are found among the H~{\sc ii} regions, including ring, shell, arc, and cometary structures.  The entire Sgr B2 complex covers an area of 
$\sim$~1$^{\prime}\times$~2$^{\prime}$, or 2.5 $\times$ 5~pc (see Fig.~\ref{fig.sb2}), which consists of more than 49 H~{\sc ii} regions, 3 of which exhibit maser activity, and a fourth region that shows maser activity but no associated H~{\sc ii} region \citep{kob89}.
The K2 region in Sgr B2 North (Fig.~\ref{fig.sb2}b) is a region of strong \water maser activity and is coincident with the dynamical center of molecular outflow (Lis et al. 1993).  DePree et al. (1995) suggest that this source is similar to source F in Sgr B2 Main (Fig. \ref{fig5.3}), another region of intense \water maser emission and bipolar molecular outflow (Lis et al. 1993; Mehringer, Goss, \& Palmer 1995a). Additionally, age determinations suggest that the Sgr B2 Main complex of H~{\sc ii} regions and the molecular outflow are related (DePree et al. 1996). From 
observations of H110$\alpha$ and H$_2$CO towards Sgr B2 North and Main, Mehringer et al. (1993, 1995b) find that Sgr B2 North is probably more distant, and suggest that two or more different events triggered star formation in each core, rather than a single global star formation event.  
\citet{tak02} find strong X-ray sources located in Sgr B2 Main associated with the H~{\sc ii} regions F3 and I, as well as a weaker source near Sgr B2 North.  The X-ray spectra and luminosities are consistent with emission arising from groups of young stellar objects (YSOs). 

The study of \water masers in high mass star forming regions has provided an important contribution to both observation and theory relating to YSO molecular outflows.  
Studies of water masers in W49 have been carried out with single dish 
antennas \citep{mor76} and Very Long Baseline Interferometry 
(VLBI) \citep{mor73,wal82,gwi92}.
Previous observations of water masers in Sgr B2 have been carried out with 
VLBI \citep{rei88} and the Nobeyama Millimeter Array \citep{kob89}.  The resolution and subsequent absolute positions from \citet{kob89} have rms errors of 1\farcs0$\times$3\farcs0 ($\alpha\times\delta$), while the relative positions have rms errors of $\sim$0\farcs3.  In order to perform a detailed comparison between the UC H~{\sc ii} regions and the maser positions, sub-arcsec resolution and positional accuracy is required.  The current observations provide an unprecedented velocity coverage of \water masers in both W49N and Sgr B2, covering a range of 870 and 160 \kms~ respectively, with angular resolutions of 0\farcs1.  At the distance of W49N \citep[11.4 kpc;][]{gwi92}, this resolution corresponds to a physical size of 0.005 pc, and for Sgr B2 at 8.5 kpc, this angular size corresponds to 0.004 pc, i.e. $<$ 1000 AU in both cases.  Additionally, the uncertainty of the relative positions of the \water masers is $\sim$0\farcs005, or a factor of at least ten better than previous observations of W49 \citep{gwi92}, and 100 times improved over previous data for Sgr B2 \citep{kob89}.  With the increased precision and velocity coverage of the current data, we can for the first time accurately compare the positions of \water masers and the UC H~{\sc ii} regions in these two sources, as well as gain a better understanding of the environments near these young stellar objects.

\section{Observations}
Observations of W49N and Sgr B2 were carried out with the VLA in both A and B configurations using the correlator in line mode 
with 2 IFs at 22 GHz.  Tab.~\ref{tab1} summarizes the parameters of these observations.  All maser observations had 63 channels per IF, with a bandwidth of 3.125 MHz, resulting in a velocity resolution of 0.66 \kms~ and a total velocity coverage of 41.6 \kms~ for each observation.  In each case, IF1 was centered on velocity 
multiples of $\pm$30 \kms~ from the \water $6_{16} \rightarrow 5_{23}$ line transition at 22.23508 GHz, which provided some overlap between observations to allow for the decreased sensitivity at the band edges.  
The velocity of 170 \kms~ was chosen as the central velocity of IF2 for both sources.  In W49N, this velocity corresponds to a bright maser feature which was used both as a positional reference point and to self-calibrate the data (see \S2.1).  In Sgr B2, this velocity is line free, which allowed us to obtain continuum images simultaneously with the line observations.
Integration times were approximately 15 minutes for each of the 29 velocity cubes in W49N, for a total of 7.3 hours on W49N, and 20 minutes for each of the 5 velocity cubes in Sgr B2, for a total of 1.7 hours on Sgr B2.  The primary phase calibrators for W49N and Sgr B2 were 1932+210 (1.44 Jy) and 1730-130 (4.73 Jy), respectively, and the primary flux density calibrator was 3C~286.  

\subsection{W49}
For W49N, we were able to self-calibrate the line cube data using a strong 
maser feature (S$>$50 Jy) at 170.7 \kms~ in IF2, observed in parallel during all 
observations. After self-calibration the typical rms noise values were 
1.0 Jy beam$^{-1}$ in a line-free channel, and 0.25 Jy beam$^{-1}$ in channels with a 
10-50 Jy beam$^{-1}$ \water maser feature. 
For the W49N 1.3cm August 1998 continuum data, IF1 was centered on v=500 \kms.  At this velocity, there are no detectable \water masers.  We were then able to use the strong maser feature at 170.7 \kms~ in IF2 to cross-calibrate the broad-band continuum at v=500 \kms~ in IF1 using the technique outlined by Reid \& Menten (1990, 1997).  This technique utilizes the phase and amplitude corrections obtained from self-calibrating the maser emission at v=170.7 \kms~ in IF2, applying these corrections to the broad-band continuum in IF1.  The resulting rms noise in the continuum was 4.3 mJy beam$^{-1}$.  Further details of the reduction of the 1.3 cm continuum data are presented by \citet{dep00}.

In order to obtain an improved alignment between the \water masers and the 1.3 cm continuum, we assumed that the position of the 170.7 \kms~ maser feature in IF2 as observed with the August 1998 continuum data was fixed.  Then all other observations of the 170.7 \kms~ maser were shifted by small amounts (typically $<$0\farcs05 in both $\alpha$ and $\delta$) in order to align the various epochs.
The shifts required to bring the IF2 170.7 \kms~ maser features into agreement with the position of the reference maser at 170.7 \kms~ in the August 1998 data were then applied to the IF1 maser line cubes using the AIPS task OGEOM.  
This method assumes negligible proper motion of the reference maser feature over the length of the observations ($\sim$1 year).  In the case of W49N, this procedure is justified since typical proper motions of these masers are on the order of a few mas per year \citep{gwi92}.  This calibration technique allowed us to obtain a relative maser alignment with respect to the 1.3 cm continuum of 0\farcs01, and a relative positional accuracy of 0\farcs005 between the masers.

\subsection{Sgr B2}
For Sgr B2, the maser lines and the continuum were observed simultaneously in order to cross-calibrate the continuum with the masers, again following the technique of Reid \& Menten (1990, 1997).  The line data was first self-calibrated, and then the corrections were applied to the line-free continuum in IF2.  Each line cube was therefore accurately aligned to its corresponding continuum.  The typical rms noise obtained for the Sgr B2 observations was similar to that of W49 (1.0 Jy beam$^{-1}$ in a line-free channel).

The centroids of the H~{\sc ii} regions in the continuum observed with the IF1 velocity cube at 40 \kms~ were found to agree with positions tabulated in \citet{gau95} to within their quoted rms errors of 0\ras01 in $\alpha$ and 0\farcs1 in $\delta$.  This continuum was used as a reference for the other data cubes.  We then shifted each observation in order to match this continuum, and applied these shifts to each corresponding IF1 line cube.  Once again, shifts were typically $<$0\farcs05 in $\alpha$ and $\delta$.  The rms error in the alignment of the \water masers and the continuum was again $\sim$0\farcs01, with a relative positioning accuracy of 0\farcs005 for the maser features in Sgr B2.

\section{Results}
\subsection{W49}
\begin{figure}
\plotone{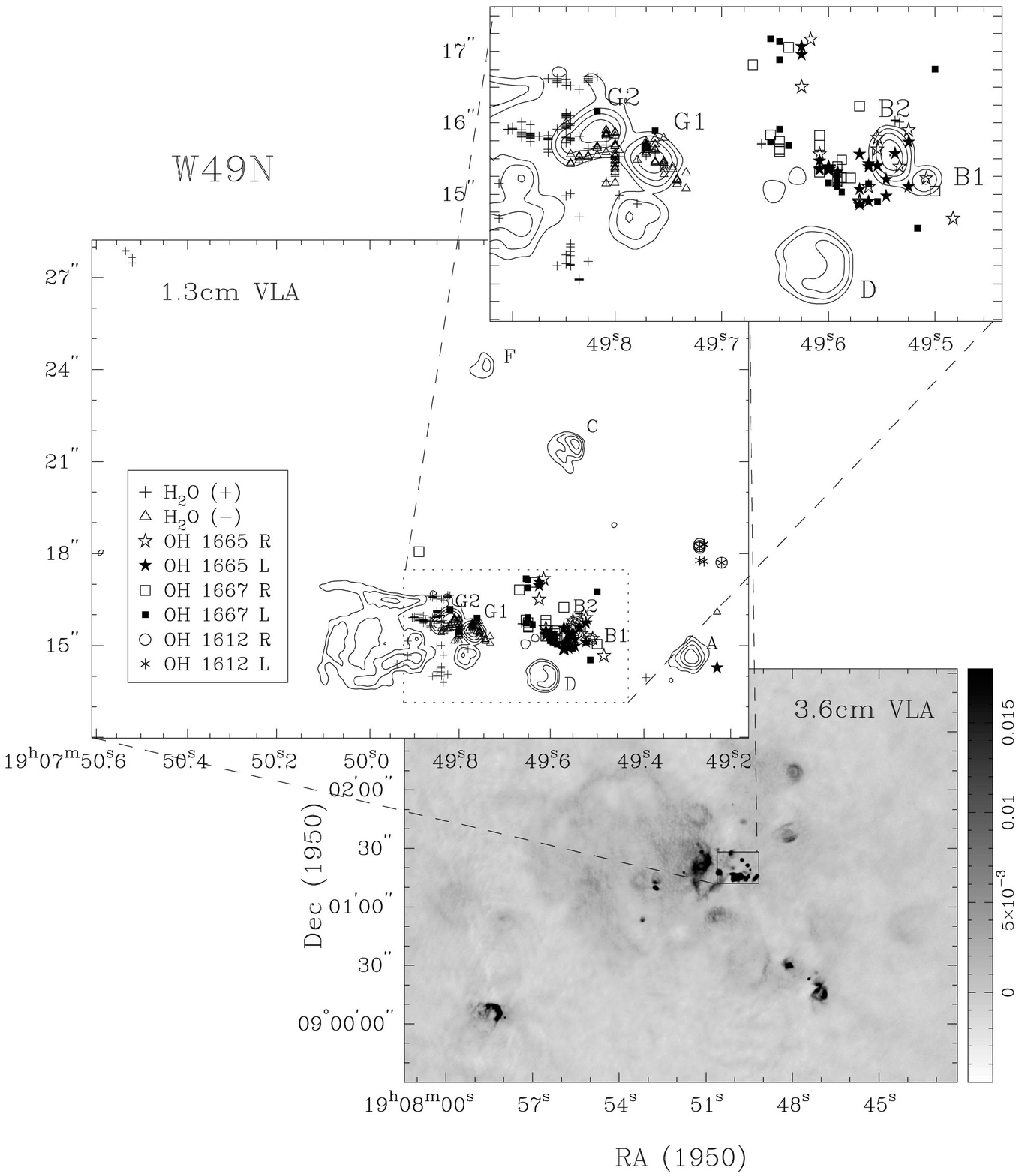}
\caption{\footnotesize Overview image of W49N continuum at 3.6 cm (from \citet{dep97}) and 1.3cm insets (from De Pree et al.~2000).  Beam size for the 3.6cm continuum is 0\farcs80 $\times$ 0\farcs78, PA = -63$^{o}$.  Beam size for the 1.3cm continuum is 0\farcs28 $\times$ 0\farcs24, PA = 8$^{o}$.  Contours are at 10, 20, 40, 80, and 160 mJy beam$^{-1}$.  \water masers (this work) and OH masers (from Argon et al.~2000) are overplotted on the 1.3cm continuum.  \water masers with positive (negative) velocities with respect to the LSR are represented by crosses (triangles), while OH masers are denoted in the key by their transition (in MHz) with an R (L) for right (left) circularly polarized masers.  
Absolute alignment between the \water and OH maser positions is limited by the absolute positional accuracy of the OH masers ($\sim$0\farcs3).  The alignment between the \water masers and the 1.3 cm continuum has an rms error of 0\farcs01 (see text for details).  
\label{fig.w49}}
\end{figure}

Over the velocity range -435 \kms~ to +435 \kms, we find 
316 \water maser components between velocities -352.1 and 375.5 \kms~ in W49N.  The positions, velocities and flux densities of these \water masers are listed in Tab.~2 and a
spectrum of the masers is shown in Fig.~\ref{fig1}.  
The same three regions of prominent maser activity noted by \citet{mor73}, \citet{wal82}, and \citet{gwi92} around H~{\sc ii} regions G1/G2 are observed, as well as the
separation between blueshifted features to the west and redshifted
features to the east of the outflow center (see Fig~\ref{fig.w49}).  
The high and low
velocity features co-exist in the region around the center of outflow,
while at greater distances from the outflow center, only high velocity
features are observed.
In addition to the masers surrounding the UC H~{\sc ii} regions G1 and G2, there is another
cluster of masers to the west around H~{\sc ii} region B2 (see Fig~\ref{fig.w49}).  This cluster consists mostly of OH masers (from Argon, Reid, \& Menten 2000), but a few redshifted \water masers ($\sim$100 \kms) are detected there as well, which suggests that these \water masers are not associated with the G1/G2 outflow, since all other \water masers west of source G1 are blueshifted.
Approximately 16\arcsec~to the northeast of source G2 there is a cluster of \water masers that is not
associated with any prominent H~{\sc ii} region.  The velocity range of these
masers is from 5.3 \kms~ to 121.3 \kms.  
There is also a group of OH masers at the
1612 MHz transition (Argon et al.~2000) approximately 3\farcs5 north of source A, that
are not associated with any detected H~{\sc ii} region.

\subsection{Sgr B2}
\begin{figure}
\plotone{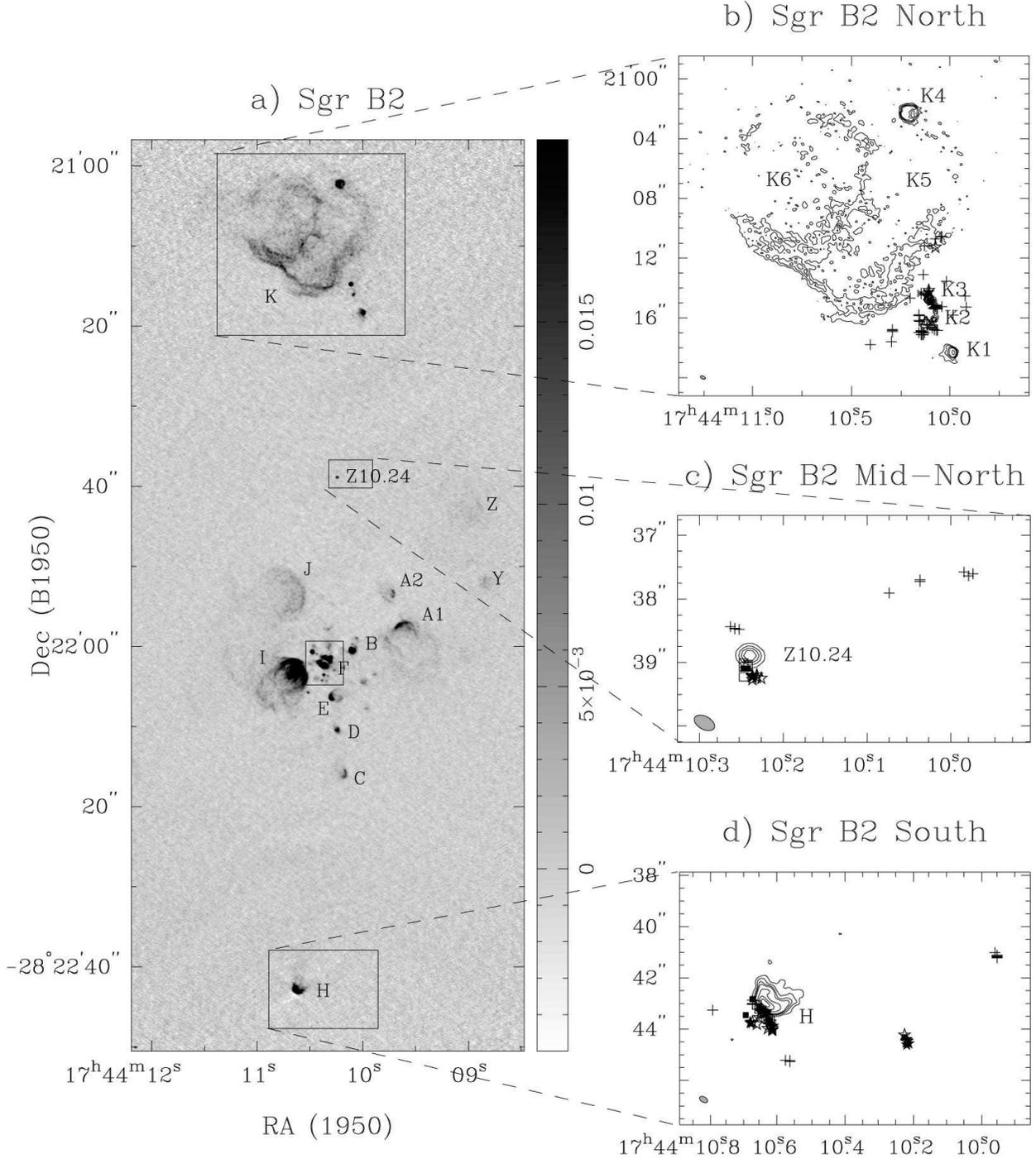}
\caption{\footnotesize Overview image of Sgr B2 continuum at 1.3 cm (from Gaume et al.~1995).  Beam size is 0\farcs35 $\times$ 0\farcs21, PA = 61$^{o}$.  Contours for images (b) through (d) are -2.5, 2.5, 5, 7.5, 10, 20, 30, 40, 50 60, 70, 80, and 90\% the peak value of 208 mJy.  The boxed region around source F is shown in greater detail in Fig. \ref{fig5.3}.  Symbols for masers plotted in figs.~(b) - (d) are similar to those plotted in Fig.~\ref{fig.w49}, with crosses for \water masers, open (filled) stars for right (left) circular polarized 1665 MHz OH masers, open (filled) boxes for right (left) circular polarized 1667 MHz OH masers, and circles (asterisks) for right (left) circular polarized 1612 MHz OH masers. The alignment error between the \water and OH masers is 0\farcs3, while the alignment error between the \water masers and the 1.3 cm continuum is 0\farcs01. \label{fig.sb2}}
\end{figure}

\begin{figure}
\plotone{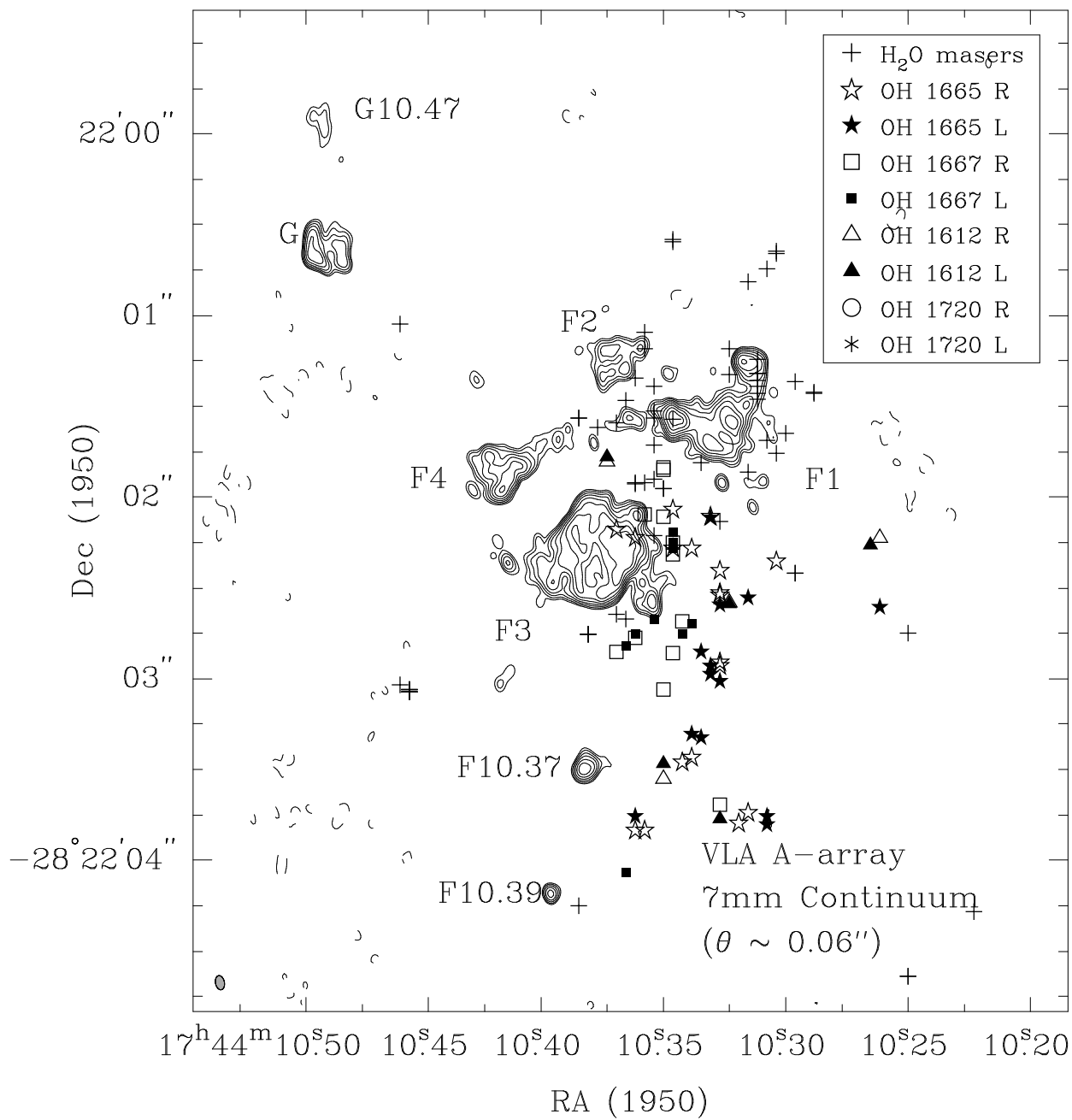}
\caption{Sgr B2 Main.  \water maser positions (this work) and OH maser positions (Argon et al. 2000) compared to the 
high resolution 7 mm continuum image of DePree et al., 1998 (beam size$=$0.\asec049 $\times$ 0.\asec079, PA=11$^{o}$).  First contour is at the 5 $\sigma$ (2.4 mJy beam$^{-1}$) level.  Successive positive contours are at 1.4, 2, 2.8, 4, 5.6, 8, 11.2, and 16 times the first contour. The alignment error between the \water and OH masers is 0\farcs3, while the alignment error between the \water masers and the 7 mm continuum is 0\farcs05. \label{fig5.3}}
\end{figure}

\begin{figure}
\plotone{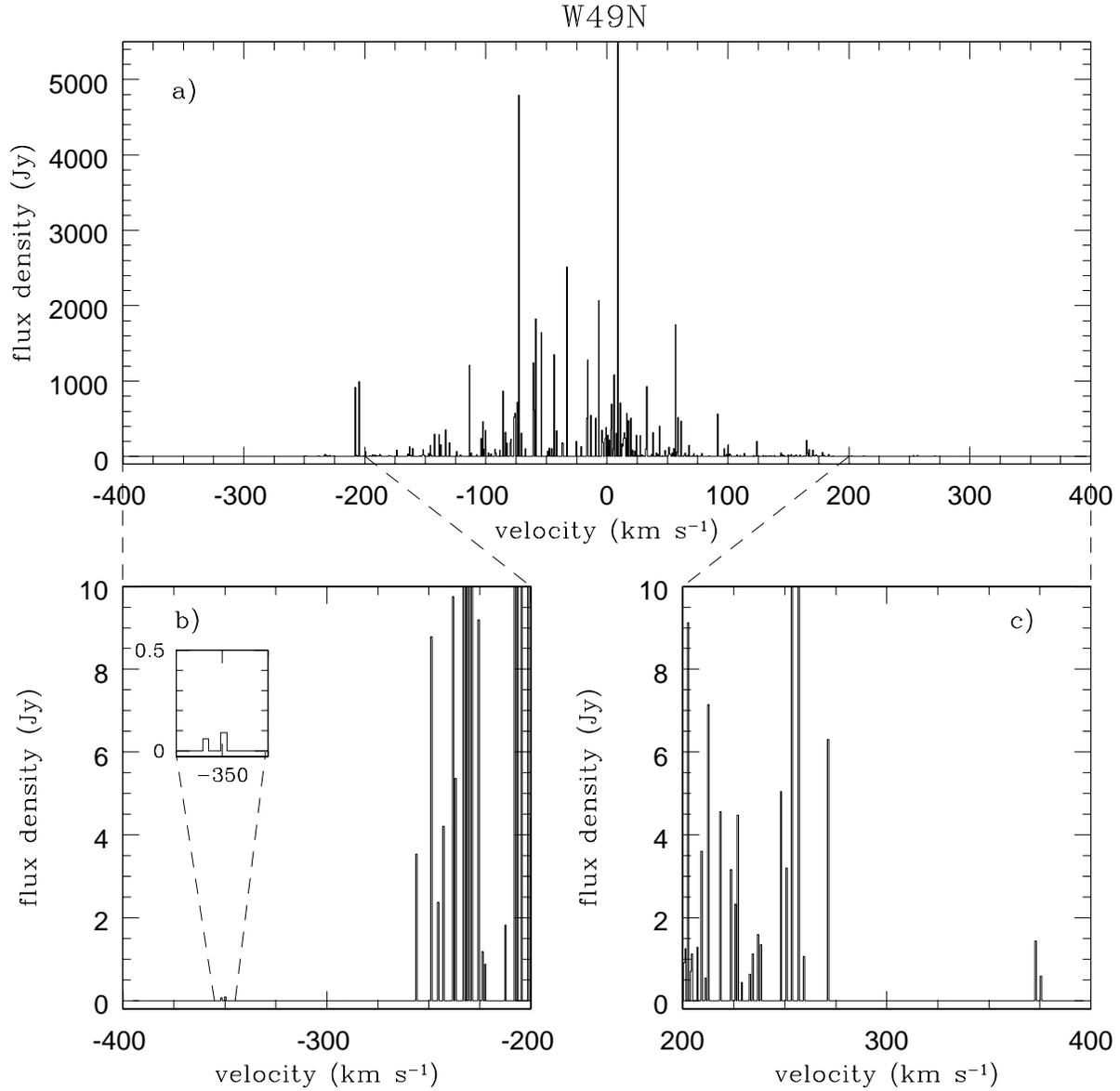}
\caption{Spectra of \water~masers in W49N observed between 1998 July and 1999 Nov.  Velocity 
is with respect to the LSR. 
The 
flux density range has been chosen so that details of the fainter, high velocity features 
are apparent.  The brightest masers are $>$ 5000~Jy beam$^{-1}$.  See Tab.~2 for a complete list of positions, velocities, and flux densities for these masers.  \label{fig1}}
\end{figure}

\begin{figure}
\plotone{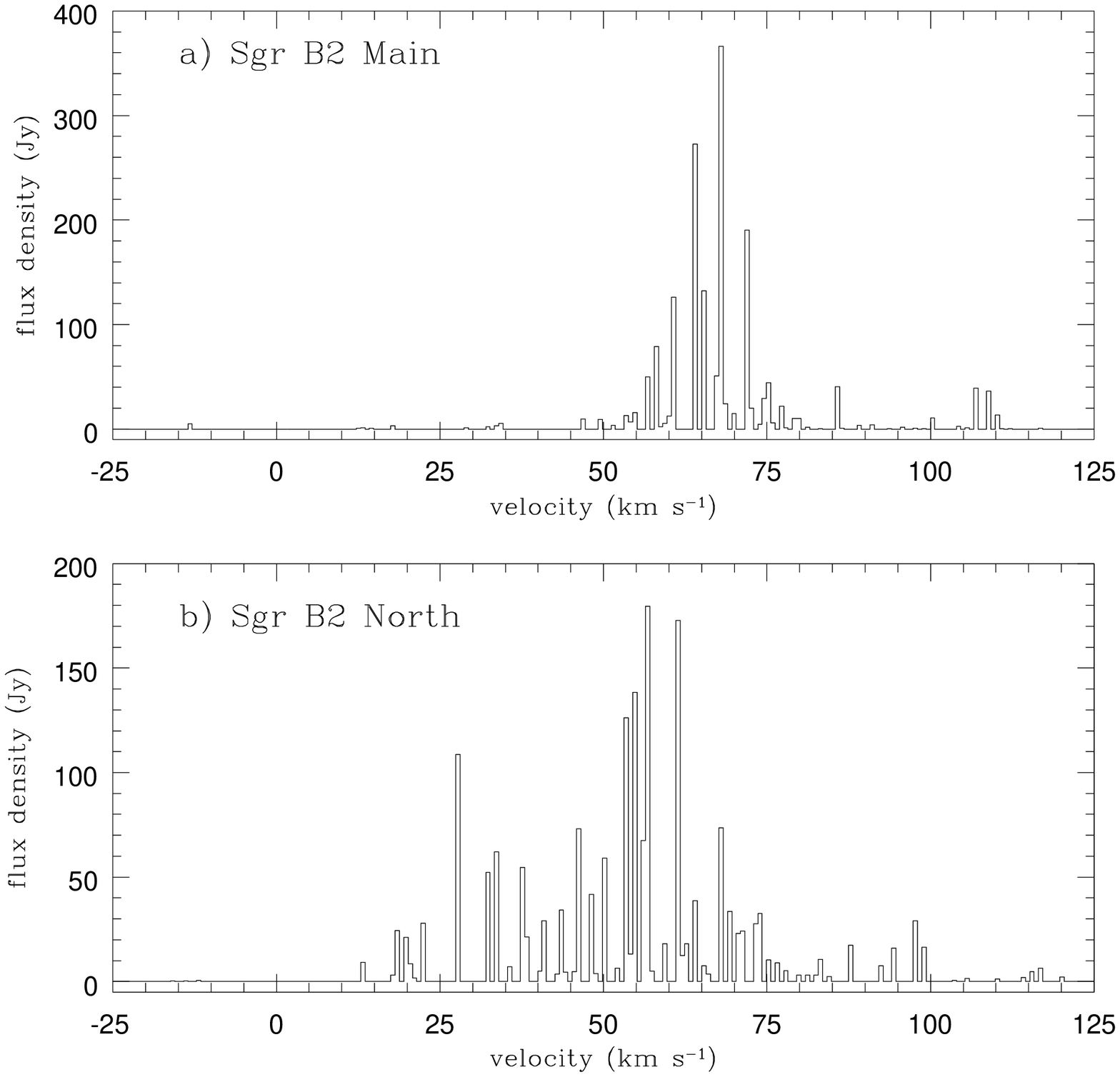}
\caption{Spectra of \water~masers in Sgr B2 Main (a) and Sgr B2 North (b) observed during 1998 June.  See Tables 3 and 4 for a complete list of positions, velocities, and flux densities for these masers. \label{fig2}}
\end{figure}

\begin{figure}
\plotone{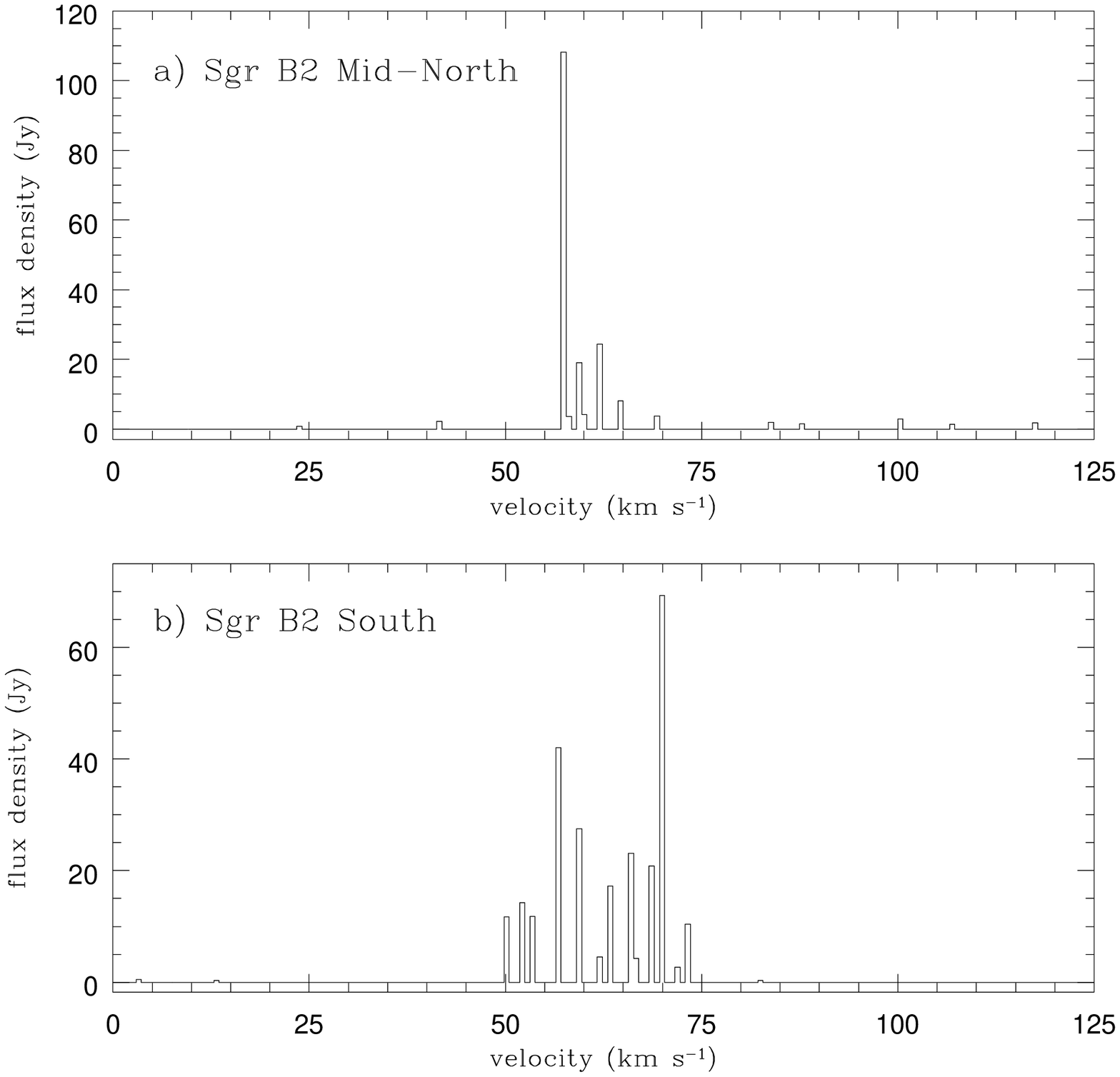}
\caption{Spectra of \water~masers in Sgr B2 Mid-North (a) and Sgr B2 South (b) observed during 1998 June.  See Tables 5 and 6 for a complete list of positions, velocities, and flux densities for these masers. \label{fig2.2}}
\end{figure}

In Sgr B2, a velocity range of -40 \kms~ to 120 \kms~ was investigated.  We detected 79 \water masers in Sgr B2 North
between -16.1 and 119.8 \kms~ (Fig.~\ref{fig.sb2}b), 14 masers in Sgr B2 Mid-North between 23.8 and 117.8 km~s$^{-1}$ (Fig.~\ref{fig.sb2}c), and 17 masers in Sgr B2 South between 3.4 and 82.8 \kms~ (Fig.~\ref{fig.sb2}d).  In Sgr
B2 Main we detected 68 masers between -13.4 and 117.1 \kms~ which are
plotted over the high resolution 7mm continuum image of \citet{dep98} in Fig.~\ref{fig5.3} along with the OH masers detected by Argon et 
al.~(2000).  Absolute positional errors for the OH masers are $\sim$0\farcs3 in $\alpha$ and $\delta$, while the absolute alignment between the \water masers and the 7 mm continuum is $\sigma\sim$0\farcs05.  Figures \ref{fig2} and \ref{fig2.2} show the entire spectrum of maser emission in Sgr B2 Main and North, and Mid-North and South, respectively, 
while Tables 3 - 6 list the positions, velocities, and flux
densities of the \water masers detected in all Sgr B2 regions.

In Sgr B2 North, the \water maser positions center around K2 (see
Fig.~\ref{fig.sb2}b), while the OH masers are clustered around K3.  
The \water masers in Sgr B2 Mid-North shown in
Fig.~\ref{fig.sb2}c are coincident with the OH masers around
the compact source labeled Z10.24 by \citet{gau95}.  A few ($<$10) high and low velocity \water masers
are visible to the west, which may indicate another region of outflow.
In Sgr B2 South, the \water and OH masers are coincident around the
front edge of the cometary source H, and have comparable velocities.
A clump of mostly low velocity \water masers to the west of
source H is detected.
Another 
clump of OH masers is detected between source H and the outlying \water masers to
the west.  

\section{Discussion}
\subsection{W49}
The observed \water maser velocity range (-352 to 375 \kms) is the largest observed for any region in the Galaxy to date, making W49N a truly exceptional \water maser source.  Due to prominent time variations in the maser emission (eg., Sullivan 1971), however, observations at other epochs may well exhibit a different velocity extreme.  Morris (1976) suggests a smaller velocity range for W49N (-256 to +271 \kms), albeit still large by comparison to other active \water maser regions in the Galaxy.
\citet{gwi92} suggest that
the \water masers in W49N are accelerating from a common origin in a bipolar outflow.
The high resolution continuum imaging of \citet{dep00} indicates that
the center of this high velocity outflow lies directly between the two
ultra-compact H~{\sc ii} regions, G1 and G2a.  

The majority of the OH masers detected by \citet{arg00} appear to be clustered
around source B2, where there is limited \water maser emission.
Additionally, \citet{gau87} find a larger cluster of OH masers
around the main region of \water maser activity, sources G1 and G2.  The
relative strength of OH and \water masers in these two H~{\sc ii} regions
may indicate different evolutionary stages, or simply different
excitation levels due to the number of stars producing UV radiation in
each region. Water masers are typically associated with the presence of
young O and B stars, but are also found to be associated with even
younger sources like hot molecular cores, and in these cases are thought
to trace the outflows associated with young stellar objects (e.g.
De~Buizer et al. 2000).  Some examples of possible hot cores indicated by the \water maser data include the region 16\asec~to the northeast of sources G1/G2 and the group of OH masers 3\farcs5 north of source A, where maser activity is observed but no associated H~{\sc ii} region is detected.  \citet{wil01} find CH$_3$CN emission at the former position which they attribute to a hot core, but do not detect any CH$_3$CN emission at the latter location.

\subsection{Sgr B2}
From Fig.~\ref{fig5.3}, we can see that the \water maser positions in Sgr B2 Main
fall almost perfectly along the outside edge of the 7mm continuum.  One interpretation of this alignment, is that entrainment may play a role in maser formation, whereby the
masers form in the dense material swept up by the stellar winds. Alternatively,
these masers may be associated with the molecular outflow of the Sgr B2 F region itself (e.g. Kuan \& Snyder 1996). 
OH masers in Sgr B2 Main are detected farther from the 7mm continuum, and there appears to be a dichotomy
between the \water and OH masers; the \water masers are found primarily
around source F1, whereas the majority of OH masers are found to the
south of F3. \citet{tor97} find a similar dichotomy in the relative positions of
OH and \water masers in W75N; the \water masers appear to be
aligned more closely along the outflow axis, while the OH masers appear
to be more distant.  They note that this arrangement is consistent with
the idea proposed by \citet{for89} in which OH masers are
formed at a later stage in the lifetime of the YSO, in the less dense
surrounding material, but they also examine the possibility that the OH
masers trace an older outflow region, distinct from that of the \water
masers.

In Sgr B2 North we see another dichotomy between \water and OH masers.  The velocities of the OH masers are lower than the velocities of the
\water masers at the same position, so it is possible that as suggested for
Sgr B2 Main, the \water and OH masers may trace different outflows, or
the OH masers may trace an older outflow event.  A few clusters of masers were discovered in areas not associated with an H~{\sc ii} region, notably the area west of Z10.24 in Sgr B2 Mid-North, and two regions west of source H in Sgr B2 South.  Each of these may indicate hot cores in various evolutionary stages and of various intensities, given the differences in the relative contribution from both \water and OH masers at each location.

\section{Conclusions}
Positions and velocities 
for 316 and 178 1.3cm \water maser components have been determined in the massive H~{\sc ii} regions, W49N and Sgr B2, respectively.  These represent the best velocity coverage
and relative maser positions to date for these sources.  
The relative positions of the continuum and maser positions have been determined to a level of 0\farcs01, while the relative positions of the \water maser components have a precision of $\sim$0\farcs005.
Both sources show high velocity outflows and strong maser 
emission regions, indicative of active star formation.  

W49N and Sgr B2 show similarities in the velocity structure of masers
around a center of outflow; close to the center, a range of
maser velocities is observed, while farther away only high velocity features are
present.  This type of structure can be explained with the model
proposed by \citet{mac92}, where the maser features lie along a shocked
shell expanding outwards from a central star or group of stars, driven
by a strong central wind.  Alternatively, \citet{gwi92} suggest 
that the masers trace the bipolar outflow itself and that a shell of shocked emission is not a necessary condition for maser formation.  

Additionally, we find that the OH and \water masers in W49N and Sgr B2
appear to trace different regions, which could be interpreted as an
evolutionary effect; the younger sources predominantly give rise to
\water masers, while the older outflows excite OH masers. In both W49N
and Sgr B2, the \water masers are found in denser regions (as traced by
molecular gas) than the OH masers, tracing out the very edge of the 7mm
continuum.  These observations support the idea that clumps of dense gas
from which the \water masers originate may be entrained by stellar winds
sweeping up the gas into an expanding shell.

Finally, several regions of \water maser activity have been detected that do not
appear to correspond with H~{\sc ii} regions (far northeast of source G in
W49N, north of source A in W49N, west of source Z10.24 in Sgr B2 Mid-North, and
two regions west of source H in Sgr B2 South).  These represent regions for further study, as they may pinpoint stars
in the earliest stages of formation, surrounded by hot cores still moving onto
the main sequence.

\acknowledgments
The authors wish to thank Crystal Brogan and the anonymous referee for helpful comments on the text, as well as Mark Reid for his advice during the initial stages of the project. EJM greatly acknowledges support from the NSF funded Research Experience for Undergraduates Program, which helped fund a part of this research.  EJM also thanks the Brandeis radio astronomy group for their hospitality at Brandeis University, where some of the data reduction was completed.  The National Radio Astronomy Observatory is a facility of the National Science Foundation operated under cooperative agreement by Associated Universities, Inc.

%\clearpage
%\input{tab1.tex}

\clearpage

\clearpage

\clearpage

\clearpage

\clearpage

\clearpage

\clearpage
\input{tab1.tex}

\clearpage
\input{tab2.tex}

\clearpage
\input{tab3.tex}

\clearpage
\input{tab4.tex}

\clearpage
\input{tab5.tex}

\clearpage
\input{tab6.tex}

\end{document}

%% file: tab1.tex
%\documentstyle[12pt,preprint]{aastex}
%\def\asec{$^{\prime\prime}$}   
%\def\arcmin{$^{\prime}$}  
%\def\and{$\&$ }
%\def\kms{km s$^{-1}$}      
%\def\pm{$\pm$}

%\begin{document} 
\begin{deluxetable}{lclccc}
\tablecolumns{6}
\tablewidth{0pc} 
\tablecaption{Observational Parameters for VLA 13 mm Observations of W49N and Sgr B2 \label{tab1}} 
\tablehead{ \colhead{Observation} & \colhead{VLA} & \colhead{IF 1 Central Vel.} &
\colhead{IF 2 Central Vel.} & \colhead{$\theta_{beam}$} & \colhead{P.A.} \\
\colhead{Date} & \colhead{Config.} & 
\colhead{(km s$^{-1}$)} & 
\colhead{(km s$^{-1}$)} & \colhead{(arcsec)} & \colhead {(deg)}}
\startdata
\cutinhead{W49N}
1998 Jul 24  &  B  &   -60,  -30,    0,  30,  60       &  170 & 0.30$\times$0.27 & 1 \\
1998 Aug 24  &  B  &  500                              &  170 & 0.28$\times$0.24 & 8 \\
1998 Sep 6   &  B  &  -150, -120,  -90,  90, 120, 150  &  170 & 0.32$\times$0.29 & 25 \\
1999 Jun 26  &  A  &  -240, -210, -180, 180, 210, 240  &  170 & 0.10$\times$0.09 & -3 \\
1999 Jul 18  &  A  &  -330, -300, -270, 270, 300, 330  &  170 & 0.11$\times$0.09 & 34 \\
1999 Nov 17  &  B  &  -420, -390, -360, 360, 390, 420  &  170 & 0.41$\times$0.28 & 44 \\
\cutinhead{Sgr B2}						 	     
1998 Jun 6   & BnA &   -20,   10,   40,  70, 100       &  170 & 0.38$\times$0.18 & 50 \\
\enddata
\end{deluxetable}
%\end{document}

%% file: tab2.tex
%\documentstyle[12pt,preprint]{aastex}
%\def\asec{$^{\prime\prime}$}   
%\def\arcmin{$^{\prime}$}  
%\def\and{$\&$ }
%\def\kms{km s$^{-1}$}      
%\def\pm{$\pm$}

%\begin{document} 
\begin{deluxetable}{ccrrr}
\tablecolumns{5}
\tablewidth{0pc} 
\tablecaption{Water Maser Positions and Velocities in W49A North \label{tab2}} 
\tablehead{ 
\colhead{$\alpha$ ($^s$)} & \colhead{$\delta$ (\asec)} & \colhead{$v_{LSR}$} & 
\colhead{Flux Density} & \colhead{Error}\\
\colhead{19$^h$07$^m$ (B1950)} & \colhead{09$^o$01\arcmin (B1950)} & 
\colhead{(km s$^{-1}$)} & \colhead{(Jy)} & \colhead{(Jy)} }
\startdata
49.904 & 15.91 &  375.5 &    0.59 &  0.20 \\
49.904 & 15.90 &  372.8 &    1.44 &  0.20 \\
49.898 & 15.92 &  271.3 &    6.30 &  0.10 \\
49.892 & 16.01 &  259.1 &    1.07 &  0.03 \\
49.893 & 15.84 &  256.5 &   17.09 &  0.04 \\
49.894 & 15.87 &  253.2 &   17.39 &  0.04 \\
49.893 & 15.84 &  250.5 &    3.20 &  0.02 \\
49.897 & 15.90 &  247.9 &    5.05 &  0.02 \\
49.889 & 15.81 &  238.0 &    1.34 &  0.02 \\
49.897 & 15.90 &  236.7 &    1.59 &  0.02 \\
49.897 & 15.91 &  234.1 &    1.12 &  0.02 \\
49.882 & 15.86 &  232.7 &    0.64 &  0.02 \\
49.900 & 15.94 &  229.1 &    0.44 &  0.03 \\
49.890 & 15.82 &  227.1 &    2.30 &  0.03 \\
49.882 & 15.86 &  226.8 &    2.18 &  0.02 \\
49.869 & 15.59 &  225.8 &    2.32 &  0.03 \\
49.849 & 16.10 &  223.8 &    3.16 &  0.03 \\
49.850 & 16.10 &  218.6 &    4.56 &  0.03 \\
49.850 & 16.10 &  212.6 &    7.14 &  0.03 \\
49.848 & 15.72 &  211.3 &    0.54 &  0.02 \\
49.887 & 15.80 &  209.3 &    3.61 &  0.03 \\
49.887 & 15.90 &  207.4 &    1.29 &  0.03 \\
49.844 & 15.75 &  204.7 &    1.12 &  0.03 \\
49.887 & 15.82 &  204.1 &    0.71 &  0.03 \\
49.852 & 16.14 &  202.7 &    9.12 &  0.03 \\
49.867 & 15.77 &  201.4 &    1.25 &  0.03 \\
49.844 & 15.75 &  200.8 &    0.91 &  0.03 \\
49.846 & 16.10 &  199.1 &    9.56 &  0.04 \\
49.887 & 15.82 &  198.1 &    1.04 &  0.03 \\
49.867 & 15.77 &  197.8 &    1.55 &  0.03 \\
49.887 & 15.82 &  196.2 &    1.73 &  0.03 \\
49.868 & 15.79 &  194.5 &    6.28 &  0.03 \\
49.844 & 15.75 &  193.2 &    4.13 &  0.03 \\
49.853 & 16.13 &  191.2 &    1.35 &  0.03 \\
49.868 & 15.79 &  189.9 &    2.24 &  0.02 \\
49.867 & 15.78 &  187.3 &    1.20 &  0.03 \\
49.844 & 15.75 &  186.6 &   11.01 &  0.03 \\
49.852 & 15.79 &  183.3 &   22.98 &  0.05 \\
49.888 & 15.82 &  183.3 &    5.87 &  0.05 \\
49.844 & 15.75 &  181.3 &    3.68 &  0.03 \\
49.841 & 15.99 &  180.7 &    2.16 &  0.03 \\
49.852 & 16.13 &  179.3 &   16.46 &  0.04 \\
49.866 & 15.76 &  178.0 &   22.12 &  0.08 \\
49.849 & 16.09 &  178.0 &   21.81 &  0.08 \\
49.845 & 15.75 &  178.0 &    5.58 &  0.08 \\
49.841 & 15.99 &  178.0 &    2.45 &  0.08 \\
49.851 & 15.80 &  177.4 &   10.17 &  0.05 \\
49.884 & 15.80 &  175.4 &    2.79 &  0.03 \\
49.879 & 15.97 &  174.7 &    1.55 &  0.03 \\
49.852 & 16.11 &  172.8 &    5.79 &  0.04 \\
49.840 & 15.96 &  172.8 &   13.21 &  0.04 \\
49.879 & 15.97 &  172.1 &    5.58 &  0.04 \\
49.852 & 16.12 &  172.1 &    5.85 &  0.04 \\
49.884 & 15.79 &  172.1 &    1.18 &  0.04 \\
49.841 & 15.98 &  170.8 &    9.8\0 &  0.1\0 \\
49.844 & 15.75 &  170.1 &   84.2\0 &  0.2\0 \\
49.855 & 16.13 &  170.1 &    4.0\0 &  0.2\0 \\
49.852 & 15.79 &  168.8 &    4.57 &  0.04 \\
49.841 & 15.98 &  166.8 &   94.7\0 &  0.2\0 \\
49.854 & 16.11 &  165.8 &   47.1\0 &  0.2\0 \\
49.844 & 15.76 &  165.2 &   85.5\0 &  0.3\0 \\
49.844 & 15.75 &  164.8 &  127.1\0 &  0.2\0 \\
49.855 & 16.13 &  162.9 &    7.42 &  0.04 \\
49.865 & 15.82 &  159.9 &   19.0\0 &  0.1\0 \\
49.849 & 16.06 &  156.6 &   25.1\0 &  0.1\0 \\
49.873 & 15.81 &  154.6 &    7.37 &  0.02 \\
49.845 & 15.76 &  152.6 &   29.70 &  0.08 \\
49.855 & 16.13 &  152.0 &   14.98 &  0.05 \\
49.835 & 13.80 &  150.0 &   21.24 &  0.06 \\
49.846 & 15.76 &  146.7 &   17.50 &  0.05 \\
49.846 & 15.76 &  145.4 &   19.92 &  0.06 \\
49.871 & 15.78 &  144.1 &    8.2\0 &  0.1\0 \\
49.835 & 13.80 &  144.1 &   38.1\0 &  0.1\0 \\
49.850 & 16.07 &  138.8 &   13.29 &  0.04 \\
49.834 & 13.80 &  138.4 &    2.3\0 &  0.1\0 \\
49.881 & 15.86 &  136.2 &    6.62 &  0.04 \\
49.900 & 15.91 &  133.8 &    7.45 &  0.03 \\
49.835 & 13.80 &  131.2 &    9.82 &  0.04 \\
49.881 & 15.82 &  129.2 &   14.16 &  0.04 \\
49.844 & 14.20 &  127.2 &    4.54 &  0.02 \\
49.834 & 13.81 &  125.3 &    9.9\0 &  0.2\0 \\
49.883 & 15.80 &  124.0 &  198.0\0 &  0.5\0 \\
50.538 & 27.85 &  121.3 &    1.42 &  0.05 \\
49.844 & 15.79 &  121.3 &   12.63 &  0.05 \\
49.835 & 13.81 &  116.7 &    4.09 &  0.02 \\
49.843 & 14.14 &  116.7 &    1.53 &  0.02 \\
49.845 & 15.78 &  113.4 &   42.6\0 &  0.1\0 \\
49.533 & 16.01 &  113.4 &    1.6\0 &  0.1\0 \\
49.846 & 15.78 &  110.1 &   13.67 &  0.04 \\
50.537 & 27.85 &  108.8 &    0.43 &  0.03 \\
49.536 & 16.02 &  107.5 &   20.42 &  0.07 \\
49.783 & 14.87 &  105.2 &    4.94 &  0.04 \\
49.837 & 15.81 &  102.5 &   22.5\0 &  0.2\0 \\
49.537 & 16.02 &  101.9 &   35.0\0 &  0.2\0 \\
49.883 & 15.74 &  100.5 &  154.8\0 &  0.5\0 \\
49.836 & 15.81 &   99.2 &   27.6\0 &  0.1\0 \\
49.536 & 16.03 &   97.2 &  102.0\0 &  0.2\0 \\
49.840 & 15.01 &   94.6 &    2.59 &  0.09 \\
49.666 & 15.70 &   93.3 &    2.14 &  0.08 \\
49.883 & 15.74 &   92.0 &  565.1\0 &  1.5\0 \\
49.667 & 15.71 &   89.3 &    1.68 &  0.03 \\
49.844 & 14.31 &   87.4 &    1.17 &  0.03 \\
49.843 & 16.52 &   85.4 &    2.87 &  0.04 \\
49.537 & 16.03 &   84.7 &    9.92 &  0.05 \\
49.803 & 15.44 &   83.4 &    5.38 &  0.04 \\
49.840 & 14.01 &   81.4 &    2.21 &  0.06 \\
49.842 & 14.27 &   80.1 &    1.88 &  0.04 \\
49.814 & 15.69 &   78.8 &   42.4\0 &  0.1\0 \\
49.849 & 16.61 &   78.1 &    6.77 &  0.08 \\
49.820 & 15.67 &   74.8 &   15.16 &  0.08 \\
49.856 & 15.91 &   72.2 &   42.3\0 &  0.3\0 \\
49.857 & 16.59 &   70.5 &    5.4\0 &  0.1\0 \\
49.830 & 16.60 &   70.5 &    5.4\0 &  0.1\0 \\
49.882 & 15.83 &   67.9 &  151.2\0 &  0.8\0 \\
49.842 & 14.02 &   66.6 &   17.1\0 &  0.2\0 \\
49.858 & 16.57 &   64.6 &   31.7\0 &  0.4\0 \\
49.824 & 15.65 &   64.6 &   14.4\0 &  0.4\0 \\
49.830 & 16.59 &   64.0 &   20.4\0 &  0.4\0 \\
49.883 & 15.78 &   61.3 &  415.1\0 &  2.0\0 \\
49.833 & 15.48 &   61.3 &   24.4\0 &  2.0\0 \\
49.840 & 14.14 &   61.3 &   29.7\0 &  2.0\0 \\
49.852 & 15.93 &   58.7 &  520.8\0 &  3.4\0 \\
49.914 & 14.66 &   58.0 &   31.9\0 &  2.8\0 \\
49.829 & 16.65 &   58.0 &   28.6\0 &  2.8\0 \\
49.936 & 14.40 &   56.7 &  148\1\0 & 10\1\0 \\
49.880 & 15.73 &   56.7 & 1410\1\0 & 10\1\0 \\
49.864 & 16.62 &   56.7 &   90.9\0 & 10.0\0 \\
49.832 & 14.06 &   56.7 &   96.5\0 & 10.0\0 \\
49.851 & 15.93 &   55.4 &   72.8\0 &  1.6\0 \\
49.849 & 14.98 &   55.4 &   35.3\0 &  1.6\0 \\
49.829 & 16.57 &   54.1 &   48.4\0 &  0.8\0 \\
49.834 & 16.47 &   52.8 &   24.2\0 &  0.4\0 \\
49.850 & 15.95 &   52.1 &   35.4\0 &  0.5\0 \\
49.854 & 16.56 &   51.4 &   34.2\0 &  0.9\0 \\
49.836 & 15.53 &   51.4 &   91.0\0 &  0.9\0 \\
49.807 & 15.53 &   50.8 &   20.7\0 &  0.7\0 \\
49.860 & 16.60 &   48.1 &   17.4\0 &  0.8\0 \\
49.837 & 16.47 &   48.1 &   62.8\0 &  0.8\0 \\
49.538 & 16.04 &   45.5 &   12.7\0 &  0.4\0 \\
49.825 & 15.64 &   43.8 &   36.8\0 &  0.6\0 \\
49.882 & 15.83 &   43.5 &  368\1\0 & 10\1\0 \\
49.858 & 16.57 &   42.5 &   27.9\0 &  0.5\0 \\
49.841 & 14.02 &   41.2 &   48.0\0 &  0.4\0 \\
49.883 & 15.81 &   38.6 &  319.0\0 &  1.5\0 \\
49.857 & 16.57 &   37.2 &   28.5\0 &  0.6\0 \\
49.865 & 15.63 &   33.3 &  164.3\0 &  2.9\0 \\
49.882 & 15.83 &   33.3 &  767.2\0 &  2.9\0 \\
49.806 & 15.49 &   32.6 &   40.9\0 &  2.6\0 \\
49.857 & 16.62 &   32.6 &   60.0\0 &  2.6\0 \\
49.537 & 16.04 &   31.3 &   11.6\0 &  0.7\0 \\
49.823 & 16.64 &   29.3 &   12.3\0 &  0.8\0 \\
49.799 & 15.53 &   29.3 &    8.87  &  0.80 \\
49.538 & 16.04 &   27.4 &   31.3\0 &  1.1\0 \\
49.883 & 15.85 &   27.4 &  245.9\0 &  1.1\0 \\
49.825 & 16.62 &   26.0 &   13.4\0 &  0.6\0 \\
49.537 & 16.03 &   24.7 &   26.3\0 &  0.8\0 \\
49.847 & 14.00 &   24.1 &   40.0\0 &  1.0\0 \\
49.882 & 15.85 &   24.1 &  218.8\0 &  1.0\0 \\
49.795 & 14.99 &   23.4 &   45.7\0 &  0.9\0 \\
50.536 & 27.87 &   23.4 &   24.5\0 &  0.9\0 \\
50.536 & 27.85 &   21.4 &   39.4\0 &  1.4\0 \\
49.775 & 15.59 &   20.8 &   18.9\0 &  1.8\0 \\
49.807 & 15.52 &   20.8 &   27.2\0 &  1.8\0 \\
49.883 & 15.84 &   20.1 &  243.1\0 &  2.3\0 \\
49.853 & 16.53 &   19.5 &  239.1\0 &  2.8\0 \\
49.912 & 15.15 &   19.5 &   27.1\0 &  2.8\0 \\
49.849 & 14.03 &   18.1 &  478\1\0 & 11\1\0 \\
49.849 & 15.92 &   16.2 &  285.5\0 &  3.8\0 \\
49.810 & 15.82 &   16.2 &  284.5\0 &  3.8\0 \\
49.770 & 15.54 &   15.5 &   85.1\0 &  2.9\0 \\
49.387 & 13.96 &   15.5 &   73.8\0 &  2.9\0 \\
49.825 & 15.49 &   14.9 &   89.3\0 &  4.1\0 \\
49.885 & 15.85 &   14.5 &  314.0\0 &  1.8\0 \\
50.522 & 27.64 &   13.8 &  237.3\0 &  3.0\0 \\
49.856 & 16.56 &   13.2 &   36.4\0 &  2.8\0 \\
49.853 & 14.15 &   13.2 &  103.6\0 &  2.8\0 \\
49.855 & 15.92 &   12.5 &  102.8\0 &  4.2\0 \\
49.837 & 13.82 &   12.5 &   69.4\0 &  4.2\0 \\
49.886 & 15.85 &   11.2 &  424.0\0 &  7.9\0 \\
49.768 & 15.60 &   11.2 &  290.3\0 &  7.9\0 \\
49.803 & 15.59 &    9.2 & 5040\1\0 & 30\1\0 \\
49.838 & 14.68 &    9.2 &  481\1\0 & 31\1\0 \\
49.857 & 14.00 &    7.9 &  313\1\0 & 24\1\0 \\
49.808 & 15.25 &    5.9 & 1080\1\0 & 10\1\0 \\
50.522 & 27.47 &    5.3 &   98.8\0 &  5.0\0 \\
49.853 & 15.90 &    3.9 &  384.0\0 &  3.1\0 \\
49.768 & 15.57 &    3.9 &  306.8\0 &  3.1\0 \\
49.801 & 15.83 &    3.3 &  362.2\0 &  2.1\0 \\
49.853 & 15.99 &    2.0 &  220.8\0 &  1.6\0 \\
49.814 & 15.18 &    0.7 &  282.0\0 &  1.4\0 \\
49.854 & 16.00 &   -0.7 &  109.6\0 &  1.8\0 \\
49.840 & 15.44 &   -0.7 &  161.1\0 &  1.8\0 \\
49.769 & 15.63 &   -0.7 &  118.0\0 &  1.8\0 \\
49.814 & 15.15 &   -1.3 &  242.5\0 &  1.7\0 \\
49.842 & 15.42 &   -3.3 &  188.2\0 &  2.1\0 \\
49.748 & 15.38 &   -3.9 &  349.9\0 &  2.6\0 \\
49.804 & 15.57 &   -6.6 & 2070\1\0 & 10\1\0 \\
49.771 & 15.70 &   -9.2 &  510.4\0 &  3.1\0 \\
49.771 & 15.69 &  -13.2 &  544.2\0 &  2.8\0 \\
49.771 & 15.62 &  -15.8 &  909\1\0 & 22\1\0 \\
49.804 & 15.16 &  -15.8 &  373\1\0 & 22\1\0 \\
49.803 & 15.31 &  -16.8 &  506.5\0 &  3.3\0 \\
49.809 & 15.83 &  -21.4 &  133.4\0 &  1.2\0 \\
49.772 & 15.66 &  -25.4 &  126.5\0 &  1.0\0 \\
49.811 & 15.86 &  -25.4 &   77.8\0 &  1.0\0 \\
49.804 & 15.35 &  -33.3 & 2520\1\0 & 10\1\0 \\
49.771 & 15.64 &  -36.6 &  178.4\0 &  1.7\0 \\
49.810 & 15.81 &  -37.2 &  182.4\0 &  1.4\0 \\
49.771 & 15.69 &  -41.8 &  339.3\0 &  3.7\0 \\
49.802 & 15.68 &  -43.8 & 1350\1\0 & 10\1\0 \\
49.773 & 15.59 &  -45.8 &  107.7\0 &  9.8\0 \\
49.814 & 15.90 &  -47.5 &  109.6\0 &  0.6\0 \\
49.805 & 15.37 &  -48.8 &   71.6\0 &  0.5\0 \\
49.803 & 15.67 &  -54.1 & 1650\1\0 & 10\1\0 \\
49.771 & 15.69 &  -58.7 & 1830\1\0 & 10\1\0 \\
49.803 & 15.68 &  -60.0 &  619.3\0 &  4.0\0 \\
49.771 & 15.69 &  -60.7 & 1240\1\0 & 10\1\0 \\
49.802 & 15.68 &  -67.2 &  106.3\0 &  1.1\0 \\
49.833 & 15.53 &  -70.9 &  308.7\0 &  3.4\0 \\
49.772 & 15.68 &  -72.5 & 4791\1\0 &  2\1\0 \\
49.833 & 15.53 &  -73.8 &  716.2\0 &  4.5\0 \\
49.802 & 15.68 &  -76.2 &  570.6\0 &  1.7\0 \\
49.771 & 15.57 &  -76.8 &  523.5\0 &  1.7\0 \\
49.832 & 15.53 &  -79.5 &  233.6\0 &  0.9\0 \\
49.812 & 15.89 &  -80.1 &  178.9\0 &  0.9\0 \\
49.805 & 15.42 &  -82.8 &  177.6\0 &  0.8\0 \\
49.832 & 15.53 &  -84.1 &  325.2\0 &  0.7\0 \\
49.832 & 15.53 &  -86.1 &  868.9\0 &  1.8\0 \\
49.767 & 15.62 &  -88.7 &   88.9\0 &  0.3\0 \\
49.799 & 15.62 &  -92.0 &   45.8\0 &  0.2\0 \\
49.770 & 15.64 &  -92.6 &   98.7\0 &  0.2\0 \\
49.831 & 15.54 &  -95.9 &   32.8\0 &  0.2\0 \\
49.805 & 15.53 &  -97.9 &   48.3\0 &  0.2\0 \\
49.772 & 15.67 & -100.5 &  349.8\0 &  0.8\0 \\
49.833 & 15.54 & -100.9 &   99.8\0 &  0.8\0 \\
49.775 & 15.74 & -102.5 &  461.7\0 &  0.9\0 \\
49.773 & 15.71 & -103.6 &  197.4\0 &  0.8\0 \\
49.832 & 15.54 & -103.8 &   40.2\0 &  0.3\0 \\
49.807 & 15.53 & -107.5 &   28.1\0 &  0.5\0 \\
49.805 & 15.31 & -112.1 &   44.7\0 &  1.3\0 \\
49.771 & 15.67 & -113.4 & 1210\1\0 &  3\1\0 \\
49.841 & 15.56 & -120.7 &   19.0\0 &  0.5\0 \\
49.805 & 15.86 & -123.9 &   67.8\0 &  0.6\0 \\
49.805 & 15.46 & -129.9 &  182.4\0 &  0.5\0 \\
49.807 & 15.54 & -133.2 &   59.6\0 &  0.3\0 \\
49.805 & 15.43 & -133.6 &   77.8\0 &  0.7\0 \\
49.772 & 15.67 & -133.6 &  219.0\0 &  0.7\0 \\
49.803 & 15.70 & -137.5 &  159.2\0 &  0.9\0 \\
49.772 & 15.66 & -138.8 &  289.1\0 &  0.7\0 \\
49.771 & 15.67 & -142.8 &  299.6\0 &  0.7\0 \\
49.801 & 15.87 & -145.4 &   22.0\0 &  0.3\0 \\
49.771 & 15.67 & -146.1 &  152.6\0 &  0.4\0 \\
49.819 & 15.65 & -147.4 &   31.6\0 &  0.2\0 \\
49.839 & 15.47 & -147.4 &   11.5\0 &  0.2\0 \\
49.817 & 15.67 & -151.3 &   22.2\0 &  0.1\0 \\
49.771 & 15.66 & -152.0 &   90.1\0 &  0.2\0 \\
49.802 & 15.82 & -152.6 &   19.8\0 &  0.2\0 \\
49.819 & 15.65 & -154.6 &    6.71 &  0.07 \\
49.772 & 15.65 & -160.3 &  103.8\0 &  0.2\0 \\
49.805 & 15.42 & -162.5 &   13.94 &  0.05 \\
49.772 & 15.65 & -162.9 &  130.8\0 &  0.2\0 \\
49.767 & 15.45 & -163.6 &    9.8\0 &  0.3\0 \\
49.804 & 15.46 & -164.2 &   33.7\0 &  0.1\0 \\
49.805 & 15.87 & -165.5 &    1.30 &  0.07 \\
49.810 & 15.59 & -166.2 &    1.36 &  0.08 \\
49.756 & 15.35 & -168.8 &    2.34 &  0.04 \\
49.767 & 15.45 & -169.5 &    3.39 &  0.04 \\
49.772 & 15.66 & -173.4 &   84.7\0 &  0.2\0 \\
49.767 & 15.45 & -174.7 &    2.60 &  0.04 \\
49.744 & 15.14 & -174.7 &    1.79 &  0.04 \\
49.815 & 15.84 & -174.7 &    1.04 &  0.04 \\
49.747 & 15.35 & -175.4 &    1.59 &  0.03 \\
49.815 & 15.83 & -178.7 &    4.61 &  0.05 \\
49.764 & 15.70 & -178.7 &    2.44 &  0.05 \\
49.784 & 15.48 & -179.3 &    4.92 &  0.05 \\
49.762 & 15.78 & -179.3 &    2.18 &  0.05 \\
49.805 & 15.48 & -180.0 &   12.51 &  0.05 \\
49.747 & 15.35 & -180.7 &    2.08 &  0.05 \\
49.765 & 15.46 & -181.3 &    2.12 &  0.04 \\
49.747 & 15.35 & -184.6 &    5.81 &  0.04 \\
49.761 & 15.79 & -185.3 &    0.75 &  0.04 \\
49.747 & 15.35 & -186.6 &    7.33 &  0.07 \\
49.763 & 15.44 & -186.6 &    4.33 &  0.07 \\
49.756 & 15.28 & -187.2 &   27.56 &  0.08 \\
49.771 & 15.68 & -188.5 &   11.18 &  0.06 \\
49.763 & 15.44 & -189.2 &    8.81 &  0.06 \\
49.754 & 15.47 & -189.9 &    4.60 &  0.06 \\
49.747 & 15.35 & -191.2 &   22.18 &  0.06 \\
49.747 & 15.35 & -193.6 &   21.96 &  0.06 \\
49.764 & 15.45 & -194.9 &    5.50 &  0.06 \\
49.770 & 15.68 & -195.5 &   12.16 &  0.07 \\
49.754 & 15.47 & -195.8 &    3.56 &  0.11 \\
49.754 & 15.46 & -196.2 &    2.89 &  0.07 \\
49.804 & 15.48 & -199.5 &   67.6\0 &  0.1\0 \\
49.747 & 15.35 & -200.1 &    4.41 &  0.10 \\
49.808 & 15.38 & -201.5 &   10.69 &  0.07 \\
49.804 & 15.48 & -204.7 &  991.1\0 &  1.9\0 \\
49.747 & 15.35 & -206.7 &   13.30 &  0.29 \\
49.803 & 15.48 & -208.0 &  917.2\0 &  1.9\0 \\
49.763 & 15.45 & -212.6 &    1.82 &  0.03 \\
49.742 & 15.19 & -222.3 &    0.88 &  0.03 \\
49.734 & 15.30 & -223.6 &    1.18 &  0.03 \\
49.818 & 15.55 & -225.5 &    0.64 &  0.03 \\
49.804 & 15.48 & -225.5 &    8.55 &  0.03 \\
49.759 & 15.44 & -228.8 &   15.55 &  0.04 \\
49.742 & 15.20 & -230.1 &   10.67 &  0.09 \\
49.805 & 15.48 & -231.5 &   15.89 &  0.05 \\
49.759 & 15.43 & -232.8 &   28.46 &  0.09 \\
49.742 & 15.21 & -236.7 &    5.36 &  0.03 \\
49.759 & 15.43 & -238.0 &    9.75 &  0.03 \\
49.754 & 15.40 & -242.6 &    4.21 &  0.02 \\
49.759 & 15.43 & -245.3 &    2.37 &  0.02 \\
49.734 & 15.08 & -248.5 &    8.79 &  0.03 \\
49.734 & 15.08 & -255.8 &    3.54 &  0.03 \\
49.231 & 16.09 & -349.5 &    0.09 &  0.01 \\
49.234 & 16.08 & -352.1 &    0.06 &  0.01 \\
\enddata
\tablecomments{Relative positional errors are $\sim$0\farcs005 in $\alpha$ and $\delta$. Absolute errors are $\sim$0\farcs05.}
\end{deluxetable}
%\end{document}

%% file: tab3.tex
%\documentstyle[12pt,preprint]{aastex}
%\def\asec{$^{\prime\prime}$}   
%\def\arcmin{$^{\prime}$}  
%\def\and{$\&$ }
%\def\kms{km s$^{-1}$}      

%\begin{document}
\begin{deluxetable}{ccrrr}
\tablecolumns{5}
\tablewidth{0pc} 
\tablecaption{Water Maser Positions and Velocities in SgrB2 Main \label{tab3}} 
\tablehead{ 
\colhead{$\alpha$ ($^s$)} & \colhead{$\delta$ (\asec)} & \colhead{$v_{LSR}$} & 
\colhead{Flux Density} & \colhead{Error}\\
\colhead{17$^h$44$^m$ (B1950)} & \colhead{-28$^o$22\arcmin (B1950)} & 
\colhead{(km s$^{-1}$)} & \colhead{(Jy)} & \colhead{(Jy)} }

\startdata 
10.361 & 01.92 & 117.1 &   0.67 & 0.02 \\
10.290 & 01.43 & 111.9 &   0.44 & 0.01 \\
10.362 & 01.92 & 111.2 &   0.47 & 0.01 \\
10.296 & 02.42 & 109.9 &  13.46 & 0.03 \\
10.457 & 03.07 & 109.2 &  36.60 & 0.05 \\
10.457 & 03.07 & 107.2 &   2.39 & 0.03 \\
10.296 & 02.42 & 106.6 &  36.74 & 0.04 \\
10.223 & 04.28 & 105.9 &   0.44 & 0.02 \\
10.362 & 01.93 & 105.9 &   0.93 & 0.02 \\
09.800 & 05.74 & 104.6 &   1.32 & 0.02 \\
10.456 & 03.07 & 104.0 &   1.17 & 0.02 \\
10.457 & 03.07 & 100.0 &  10.81 & 0.03 \\
09.802 & 05.73 &  99.3 &   0.54 & 0.03 \\
10.457 & 03.06 &  97.4 &   0.77 & 0.04 \\
10.353 & 01.95 &  95.4 &   1.81 & 0.02 \\
10.250 & 04.64 &  93.4 &   0.52 & 0.02 \\
10.313 & 01.32 &  90.8 &   4.32 & 0.02 \\
10.352 & 01.95 &  88.8 &   1.64 & 0.02 \\
10.386 & 01.56 &  88.8 &   1.26 & 0.02 \\
10.250 & 04.64 &  88.8 &   0.96 & 0.02 \\
10.314 & 01.32 &  86.2 &   0.95 & 0.03 \\
10.385 & 01.56 &  85.5 &  40.63 & 0.04 \\
10.354 & 01.71 &  82.9 &   0.56 & 0.02 \\
10.315 & 00.81 &  81.2 &   1.88 & 0.02 \\
10.382 & 02.76 &  79.9 &  10.07 & 0.03 \\
10.354 & 01.90 &  79.2 &  10.06 & 0.03 \\
10.460 & 03.03 &  77.9 &   1.45 & 0.04 \\
10.347 & 01.57 &  77.2 &   5.60 & 0.04 \\
10.382 & 02.76 &  77.2 &  16.16 & 0.04 \\
10.346 & 00.59 &  75.9 &   6.12 & 0.07 \\
10.382 & 02.76 &  75.3 &  36.47 & 0.08 \\
10.295 & 01.37 &  75.3 &   7.81 & 0.08 \\
10.336 & 01.81 &  74.6 &  29.41 & 0.06 \\
10.360 & 01.18 &  74.0 &   4.44 & 0.06 \\
10.388 & 04.25 &  72.6 &  19.85 & 0.08 \\
10.457 & 03.06 &  72.0 &   8.3\0 & 0.2\0 \\
10.324 & 01.33 &  72.0 &  31.9\0 & 0.2\0 \\
10.303 & 01.76 &  72.0 &   7.1\0 & 0.2\0 \\
10.371 & 01.59 &  72.0 & 143.0\0 & 0.2\0 \\
10.290 & 01.42 &  70.0 &  14.6\0 & 0.4\0 \\
10.458 & 03.06 &  68.7 &  10.6\0 & 0.2\0 \\
10.310 & 01.69 &  68.7 &  13.8\0 & 0.2\0 \\
10.377 & 01.62 &  68.0 & 366.1\0 & 0.4\0 \\
10.312 & 01.39 &  67.4 &  50.9\0 & 0.3\0 \\
10.301 & 01.65 &  65.4 & 132.4\0 & 0.2\0 \\
10.313 & 01.46 &  64.1 & 272.7\0 & 0.3\0 \\
10.314 & 01.36 &  60.8 & 126.3\0 & 0.3\0 \\
10.360 & 01.09 &  60.1 &  12.4\0 & 0.1\0 \\
10.315 & 01.86 &  59.5 &   5.35 & 0.10 \\
10.348 & 00.58 &  58.8 &   2.11 & 0.09 \\
10.313 & 01.43 &  58.1 &  78.8\0 & 0.1\0 \\
10.327 & 02.13 &  56.8 &   2.7\0 & 0.2\0 \\
10.363 & 01.34 &  56.5 &  47.4\0 & 0.3\0 \\
10.303 & 00.65 &  54.5 &  15.8\0 & 0.2\0 \\
10.355 & 01.56 &  53.8 &   7.1\0 & 0.2\0 \\
10.303 & 00.66 &  53.2 &  13.1\0 & 0.2\0 \\
10.356 & 01.52 &  51.2 &   3.45 & 0.13 \\
10.462 & 01.05 &  49.2 &   9.26 & 0.10 \\
10.357 & 01.39 &  46.6 &   9.65 & 0.12 \\
10.354 & 02.21 &  34.1 &   5.64 & 0.08 \\
10.249 & 02.75 &  33.4 &   3.33 & 0.10 \\
10.312 & 01.24 &  32.1 &   2.40 & 0.08 \\
10.323 & 01.18 &  29.1 &   1.41 & 0.12 \\
10.369 & 02.67 &  17.9 &   3.38 & 0.33 \\
10.308 & 00.74 &  14.6 &   0.90 & 0.02 \\
10.371 & 02.65 &  13.3 &   1.47 & 0.03 \\
10.366 & 01.47 &  12.6 &   0.83 & 0.02 \\
09.758 & 00.64 & -13.4 &   4.95 & 0.02 \\
\enddata
\tablecomments{Relative positional errors are $\sim$0\farcs005 in $\alpha$ and $\delta$. Absolute errors are $\sim$0\farcs05.}
\end{deluxetable}
%\end{document}

%% file: tab4.tex
%\documentstyle[12pt,preprint]{aastex}
%\def\asec{$^{\prime\prime}$}   
%\def\arcmin{$^{\prime}$}  
%\def\and{$\&$ }
%\def\kms{km s$^{-1}$}      

%\begin{document}
\begin{deluxetable}{ccrrr}
\tablecolumns{5}
\tablewidth{0pc} 
\tablecaption{Water Maser Positions and Velocities in SgrB2 North \label{tab4}} 
\tablehead{ 
\colhead{$\alpha$ ($^s$)} & \colhead{$\delta$ (\asec)} & \colhead{$v_{LSR}$} & 
\colhead{Flux Density} & \colhead{Error}\\
\colhead{17$^h$44$^m$ (B1950)} & \colhead{-28$^o$21\arcmin (B1950)} & 
\colhead{(km s$^{-1}$)} & \colhead{(Jy)} & \colhead{(Jy)} }
\startdata
10.044 & 15.24 & 119.8 &   2.19 & 0.03 \\
10.064 & 16.83 & 116.5 &   6.51 & 0.04 \\
09.924 & 14.54 & 115.8 &   4.85 & 0.04 \\
10.071 & 15.21 & 114.5 &   2.05 & 0.02 \\
10.087 & 15.34 & 110.5 &   1.27 & 0.03 \\
10.162 & 15.84 & 105.3 &   1.60 & 0.03 \\
10.089 & 16.67 & 103.3 &   0.63 & 0.02 \\
10.089 & 16.68 &  99.3 &  16.37 & 0.07 \\
10.089 & 16.68 &  98.0 &  19.91 & 0.08 \\
10.161 & 15.85 &  97.4 &   9.14 & 0.08 \\
10.083 & 15.30 &  94.7 &   6.14 & 0.04 \\
10.073 & 15.24 &  94.1 &   9.91 & 0.05 \\
10.089 & 16.68 &  92.1 &   7.57 & 0.04 \\
10.067 & 15.20 &  87.5 &  17.51 & 0.07 \\
10.069 & 15.18 &  84.5 &   1.81 & 0.03 \\
10.118 & 16.45 &  84.2 &   0.54 & 0.03 \\
10.155 & 16.99 &  83.2 &   8.37 & 0.05 \\
10.082 & 15.39 &  83.2 &   2.25 & 0.05 \\
10.071 & 15.14 &  82.5 &   3.09 & 0.04 \\
09.984 & 15.81 &  81.2 &   3.07 & 0.03 \\
09.917 & 15.28 &  79.9 &   3.22 & 0.04 \\
10.070 & 15.22 &  77.9 &   5.28 & 0.05 \\
10.116 & 16.44 &  76.6 &   3.56 & 0.07 \\
10.092 & 16.46 &  76.6 &   5.43 & 0.07 \\
10.131 & 11.18 &  75.3 &   1.98 & 0.10 \\
10.039 & 10.54 &  75.3 &   8.40 & 0.10 \\
10.067 & 15.19 &  74.0 &   3.13 & 0.11 \\
10.131 & 17.01 &  74.0 &  17.0\0 & 0.1\0 \\
10.039 & 10.56 &  74.0 &  12.4\0 & 0.1\0 \\
10.405 & 17.79 &  73.3 &  27.6\0 & 0.1\0 \\
10.294 & 16.82 &  71.3 &  24.3\0 & 0.2\0 \\
10.405 & 17.79 &  70.7 &  23.0\0 & 0.2\0 \\
10.146 & 16.95 &  69.3 &  33.5\0 & 0.4\0 \\
10.293 & 16.82 &  68.0 &  73.6\0 & 0.5\0 \\
10.135 & 13.10 &  66.0 &   3.6\0 & 0.3\0 \\
10.296 & 17.60 &  65.4 &   3.3\0 & 0.3\0 \\
09.467 & 12.36 &  65.4 &   4.4\0 & 0.3\0 \\
10.017 & 13.55 &  64.1 &  38.7\0 & 0.3\0 \\
10.293 & 16.83 &  62.8 &  18.1\0 & 0.3\0 \\
10.143 & 16.94 &  62.1 &  12.5\0 & 0.4\0 \\
10.065 & 15.31 &  61.4 & 172.9\0 & 0.7\0 \\
10.293 & 16.79 &  59.5 &   2.2\0 & 0.1\0 \\
10.140 & 16.94 &  59.5 &  16.0\0 & 0.1\0 \\
10.063 & 15.36 &  57.1 &   5.1\0 & 0.4\0 \\
10.150 & 16.86 &  56.5 & 160.2\0 & 0.6\0 \\
10.074 & 10.70 &  56.5 &  19.2\0 & 0.6\0 \\
10.203 & 14.67 &  55.8 &  67.5\0 & 0.5\0 \\
10.146 & 16.90 &  55.1 & 131.9\0 & 0.5\0 \\
10.155 & 15.80 &  54.5 &   6.5\0 & 0.4\0 \\
10.157 & 16.19 &  53.8 &   7.1\0 & 0.3\0 \\
10.292 & 16.92 &  53.8 &   6.0\0 & 0.3\0 \\
10.140 & 16.97 &  53.2 & 126.2\0 & 0.5\0 \\
10.151 & 16.24 &  51.9 &   6.5\0 & 0.3\0 \\
10.068 & 15.40 &  49.9 &  59.1\0 & 0.3\0 \\
10.161 & 15.83 &  48.6 &   3.9\0 & 0.2\0 \\
10.140 & 16.97 &  47.9 &  41.7\0 & 0.2\0 \\
10.068 & 15.40 &  45.9 &  73.2\0 & 0.3\0 \\
10.158 & 16.18 &  45.3 &   4.7\0 & 0.1\0 \\
10.114 & 16.66 &  44.0 &   4.6\0 & 0.1\0 \\
10.138 & 16.98 &  43.3 &  34.2\0 & 0.1\0 \\
10.093 & 16.63 &  42.6 &   3.54 & 0.08 \\
10.139 & 16.98 &  40.7 &  29.1\0 & 0.1\0 \\
10.114 & 16.65 &  40.0 &   5.10 & 0.08 \\
10.092 & 16.62 &  38.0 &  21.5\0 & 0.1\0 \\
10.136 & 17.11 &  37.4 &  54.5\0 & 0.2\0 \\
10.072 & 16.83 &  35.4 &   7.19 & 0.10 \\
10.137 & 17.15 &  33.4 &  62.2\0 & 0.2\0 \\
10.136 & 17.15 &  32.1 &  52.2\0 & 0.2\0 \\
10.142 & 17.12 &  27.8 & 108.6\0 & 0.4\0 \\
10.143 & 17.13 &  22.5 &  27.9\0 & 0.1\0 \\
10.047 & 10.57 &  21.2 &   1.63 & 0.05 \\
10.078 & 16.78 &  20.5 &   8.59 & 0.08 \\
10.143 & 17.13 &  19.9 &  21.16 & 0.08 \\
10.142 & 17.11 &  18.6 &  24.33 & 0.08 \\
10.100 & 16.51 &  17.9 &   3.11 & 0.06 \\
10.134 & 16.13 &  13.3 &   9.14 & 0.04 \\
10.124 & 16.28 & -12.1 &   0.49 & 0.02 \\
10.122 & 16.29 & -14.1 &   0.42 & 0.03 \\
10.092 & 16.57 & -16.1 &   0.38 & 0.03 \\
\enddata
\tablecomments{Relative positional errors are $\sim$0\farcs005 in $\alpha$ and $\delta$. Absolute errors are $\sim$0\farcs05.}
\end{deluxetable}
%\end{document}

%% file: tab5.tex
%\documentstyle[12pt,preprint]{aastex}
%\def\asec{$^{\prime\prime}$}   
%\def\arcmin{$^{\prime}$}  
%\def\and{$\&$ }
%\def\kms{km s$^{-1}$}      

%\begin{document}
\begin{deluxetable}{ccrrr}
\tablecolumns{5}
\tablewidth{0pc} 
\tablecaption{Water Maser Positions and Velocities in SgrB2 Mid-North \label{tab5}} 
\tablehead{ 
\colhead{$\alpha$ ($^s$)} & \colhead{$\delta$ (\asec)} & \colhead{$v_{LSR}$} & 
\colhead{Flux Density} & \colhead{Error}\\
\colhead{17$^h$44$^m$ (B1950)} & \colhead{-28$^o$21\arcmin (B1950)} & 
\colhead{(km s$^{-1}$)} & \colhead{(Jy)} & \colhead{(Jy)} }
\startdata
09.972 & 37.60 & 117.8 &   1.81 & 0.02 \\
09.972 & 37.60 & 106.6 &   1.36 & 0.04 \\
09.972 & 37.60 & 100.0 &   2.86 & 0.03 \\
10.259 & 38.47 &  88.1 &   1.55 & 0.03 \\
10.259 & 38.46 &  83.8 &   1.87 & 0.03 \\
09.971 & 37.60 &  69.3 &   3.8\0 & 0.3\0 \\
10.263 & 38.44 &  64.7 &   8.1\0 & 0.2\0 \\
10.263 & 38.44 &  62.1 &  24.4\0 & 0.3\0 \\
10.038 & 37.70 &  60.1 &   4.2\0 & 0.2\0 \\
10.074 & 37.91 &  59.5 &  19.0\0 & 0.1\0 \\
10.253 & 38.48 &  57.8 &   3.7\0 & 0.2\0 \\
10.035 & 37.73 &  57.5 & 108.1\0 & 0.3\0 \\
09.977 & 37.64 &  41.3 &   2.22 & 0.05 \\
09.986 & 37.58 &  23.8 &   0.78 & 0.04 \\
\enddata
\tablecomments{Relative positional errors are $\sim$0\farcs005 in $\alpha$ and $\delta$. Absolute errors are $\sim$0\farcs05.}
\end{deluxetable}
%\end{document}

%% file: tab6.tex
%\documentstyle[12pt,preprint]{aastex}
%\def\asec{$^{\prime\prime}$}   
%\def\arcmin{$^{\prime}$}  
%\def\and{$\&$ }
%\def\kms{km s$^{-1}$}      

%\begin{document}
\begin{deluxetable}{ccrrr}
%\begin{table}
\tablecolumns{5}
\tablewidth{0pc} 
\tablecaption{Water Maser Positions and Velocities in SgrB2 South \label{tab6}} 
\tablehead{ 
\colhead{$\alpha$ ($^s$)} & \colhead{$\delta$ (\asec)} & \colhead{$v_{LSR}$} & 
\colhead{Flux Density} & \colhead{Error}\\
\colhead{17$^h$44$^m$ (B1950)} & \colhead{-28$^o$22\arcmin (B1950)} & 
\colhead{(km s$^{-1}$)} & \colhead{(Jy)} & \colhead{(Jy)} }
\startdata
09.949 & 41.20 &  82.8 &  0.29 & 0.02 \\
09.951 & 41.18 &  73.3 & 10.4\0 & 0.1\0 \\
10.671 & 42.87 &  72.0 &  2.7\0 & 0.2\0 \\
10.580 & 45.22 &  70.0 & 69.3\0 & 0.3\0 \\
10.672 & 43.02 &  68.7 & 20.8\0 & 0.2\0 \\
10.793 & 43.26 &  66.7 &  4.3\0 & 0.2\0 \\
10.672 & 43.01 &  66.0 & 20.0\0 & 0.2\0 \\
10.567 & 45.27 &  66.0 &  3.1\0 & 0.2\0 \\
10.673 & 43.03 &  63.4 & 17.2\0 & 0.2\0 \\
10.568 & 45.23 &  62.1 &  4.5\0 & 0.2\0 \\
09.949 & 41.11 &  59.5 & 27.5\0 & 0.1\0 \\
09.953 & 41.22 &  56.5 & 42.0\0 & 0.3\0 \\
09.950 & 41.17 &  53.2 & 11.8\0 & 0.3\0 \\
09.951 & 41.16 &  51.9 & 14.2\0 & 0.2\0 \\
09.951 & 41.16 &  49.9 & 11.7\0 & 0.2\0 \\
09.958 & 41.01 &  13.3 &  0.37 & 0.03 \\
09.959 & 41.01 &   3.4 &  0.52 & 0.03 \\
\enddata
\tablecomments{Relative positional errors are $\sim$0\farcs005 in $\alpha$ and $\delta$. Absolute errors are $\sim$0\farcs05.}
\end{deluxetable}
%\end{document}

%% file: ms.bbl
\clearpage
\begin{thebibliography}{}
\bibitem[Alves \& Homeier(2003)]{alv03} Alves, J., \& Homeier, N. 2003, 
    \apjl, 589, L45
\bibitem[Argon et al.(2000)]{arg00} Argon, A. L., Reid, M. J., \& 
    Menten, K. M. 2000, \apjs, 129, 159
\bibitem[Conti \& Blum(2002)]{con02} Conti, P. S., \& Blum, R. D. 2002, 
    \apj, 564, 827
%\bibitem[Cornwell \& Fomalont (1998)]{cor98} Cornwell, T. \& Fomalont, E. B. 1998, in \emph{Synthesis Imaging in Radio Astronomy II}, eds. G. B. Taylor, C. L. Carilli, \& R. A. Perley (San Francisco: ASP), 187
\bibitem[De~Buizer et al.(2000)]{deb00} De Buizer, J. M., Pi\~na, R. K., 
	\& Telesco, C. M. 2000, \apjs, 130, 437
\bibitem[DePree et al.(1995)]{dep95} DePree, C. G., Gaume, R. A., Goss, 
    W. M., \& Claussen, M. J. 1995, \apj, 451, 284
\bibitem[DePree et al.(1996)]{dep96} DePree, C. G., Gaume, R. A., Goss, 
    W. M., \& Claussen, M. J. 1996, \apj, 464, 788
\bibitem[DePree et al.(1997)]{dep97} DePree, C. G., Mehringer, D. M., \& 
    Goss, W. M. 1997, \apj, 482, 307
\bibitem[DePree et al.(1998)]{dep98} DePree, C. G., Goss, W. M., \& Gaume, 
    R. A. 1998, \apj, 500, 847
\bibitem[DePree et al.(2000)]{dep00} DePree, C. G., Wilner, D. J., Goss, W. 
    M., Welch, W. J., \& McGrath, E. J. 2000, \apj, 540, 308
%\bibitem[Elitzur(1992)]{eli92} Elitzur, M. 1992, \araa, 30, 75
\bibitem[Elitzur(1995)]{eli95} Elitzur, M. 1995, in RevMexAA Ser. de Conf. 1, 
    Circumstellar Disks, Outflows and Star Formation, eds. S. Lizano, \& 
    J. M. Torrelles (Cozumel, Mexico: RevMexAA), 85
\bibitem[Garay \& Lizano(1999)]{gar99} Garay, G., \& Lizano, S. 1999, \pasp, 
    111, 1049
\bibitem[Gaume et al.(1995)]{gau95} Gaume, R. A., Claussen, M. J., DePree, 
    C. G., Goss, W. M., \& Mehringer, D. M. 1995, \apj, 449, 663
\bibitem[Gaume \& Mutel(1987)]{gau87} Gaume, R. A., \& Mutel, R. L. 1987, 
    \apjs, 65, 193
%\bibitem[Gundermann(1965)]{gun65} Gundermann, E. 1965, Ph. D. thesis, 
%    Harvard University, Cambridge, Massachusetts
\bibitem[Gwinn et al.(1992)]{gwi92} Gwinn, C. R., Moran, J. M., \& Reid, 
    M. J. 1992, \apj, 393, 149
\bibitem[Forster \& Caswell(1989)]{for89} Forster, J. R., \& Caswell, J. L. 
    1989, \aap, 213, 339
\bibitem[Kobayashi et al.(1989)]{kob89} Kobayashi, H., Ishiguro, M., 
    Chikada, Y., Ukita, N., Morita, K.-I., Okumura, S. K., Kasuga, T., \& 
    Kawabe, R. 1989, \pasj, 41, 141
\bibitem[Kuan \& Snyder(1996)]{kua96} Kuan, Y.-J., \& Snyder, L. E. 1996, 
    \apj, 470, 981
\bibitem[Lis et al.(1993)]{lis93} Lis, D. C., Goldsmith, P. F., Carlstrom, 
    J. E., \& Scoville, N. Z. 1993, \apj, 402, 238
\bibitem[Mac Low \& Elitzur(1992)]{mac92} Mac Low, M.-M., \& Elitzur, M. 
    1992, \apj, 393, 33
\bibitem[Mehringer et al.(1993)]{meh93} Mehringer, D. M., Palmer, P., 
    Goss, W. M., \& Yusef-Zadeh, F. 1993, \apj, 412, 684
\bibitem[Mehringer et al.(1995a)]{meh95a} Mehringer, D. M., Goss, W. M., \& 
    Palmer, P. 1995a, \apj, 452, 304
\bibitem[Mehringer et al.(1995b)]{meh95b} Mehringer, D. M., Palmer, P., \& 
    Goss, W. M. 1995b, \apjs, 97, 497
\bibitem[Mezger et al.(1967)]{mez67} Mezger, P. G., Schraml, J., \& Terzian, 
    Y. 1967, \apj, 150, 807
\bibitem[Moran et al.(1973)]{mor73} Moran J. M., Papadopoulos, G. D., Burke, 
    B. F., Lo, K. Y., Schwartz, P. R., \& Thacker, D. L. 1973, \apj, 185, 535
\bibitem[Morris(1976)]{mor76} Morris, M. 1976, \apj, 210, 100
\bibitem[Reid \& Menten(1990)]{rei90} Reid, M. J., \& Menten, K. M. 1990, 
    \apjl, 360, L51
\bibitem[Reid \& Menten(1997)]{rei97} Reid, M. J., \& Menten, K. M. 1997, 
    \apj, 476, 327
%\bibitem[Reid and Moran(1981)]{rei81} Reid, M. J., \& Moran, J. M.,  1981, 
%    \araa, 19, 231
%\bibitem[Reid and Moran(1988)]{rei88} Reid, M. J., \& Moran, J. M.,  1988, 
%    in \emph{Galactic and Extragalactic Radio 
%    Astronomy}, eds. G. L. Verschuur, K. I. Kellerman, and E. Bouton 
%    (New York:Springer-Verlag), p?
\bibitem[Reid et al.(1988)]{rei88} Reid, M. J., Schneps, M. H., Moran, J. M., 
    Gwinn, C. R., Genzel, R., Downes, D., \& R\"onn\"ang, B. 1988, \apj, 330, 
    809
%\bibitem[Scheffler and Els\"asser(1987)]{she87} Scheffler, H., \& 
%    Els\"asser, H. 1987, \emph{Physics of the Galaxy and Interstellar Matter},
%    Springer-Verlag, New York
\bibitem[Sullivan(1971)]{sul71} Sullivan, W. T., III 1971, \apj, 166, 321
\bibitem[Takagi et al.(2002)]{tak02} Takagi, S., Murakami, H., \& Koyama, K. 
    2002, \apj, 573, 275
\bibitem[Torrelles et al.(1997)]{tor97} Torrelles, J. M., G\'omez, J. F., 
    Rodr\'iguez, L. F., Ho, P. T. P., Curiel, S., \& V\'azquez, R. 1997, 
    \apj, 489, 744
%\bibitem[Vogel et al.(1987)]{vog87} Vogel, S. N., Genzel, R., \& Palmer, P. 
%    1987, \apj, 316, 243
\bibitem[Walker et al.(1982)]{wal82} Walker, R. C., Matsakis, D. N., \&
    Garcia-Barreto, J. A. 1982, \apj, 255, 128
%\bibitem[Weaver et al.(1965)]{wea65} Weaver, H., Williams, D. R. W., Dieter, 
%    N. H., \& Lum, W. T. 1965, \nat, 208, 29
\bibitem[Wilner et al.(2001)]{wil01} Wilner, D. J., De Pree, C. G., Welch, W. J., \& Goss, W. M. 2001, \apjl, 550, L81
\end{thebibliography}
